\documentclass[12pt,twoside]{article}
\usepackage[T1]{fontenc}
\usepackage[latin1]{inputenc}
\usepackage{times}
\usepackage{graphics}
\usepackage{a4wide}

\begin{document}

\section*{Elementary Collisions with HADES\footnote{Presentation given on the 11th International Conference on
Meson-Nucleon Physics and
the Structure of the Nucleon,
September 10-14, 2007, Juelich, Germany
}}

\thispagestyle{empty}

\begin{raggedright}


I.~Fr\"{o}hlich$^{7}$, J.~Pietraszko$^{4}$, G.~Agakishiev$^{8}$, C.~Agodi$^{1}$, A.~Balanda$^{3}$, G.~Bellia$^{1}$,
D.~Belver$^{15}$, A.~Belyaev$^{6}$, A.~Blanco$^{2}$, M.~B\"{o}hmer$^{11}$
% ...

\begin{center} (HADES collaboration)
\end{center}


{$^{1}$}Istituto Nazionale di Fisica Nucleare - Laboratori Nazionali del Sud, 95125~Catania, Italy\\
{$^{2}$LIP-Laborat\'{o}rio de Instrumenta\c{c}\~{a}o e F\'{\i}sica Experimental de Part\'{\i}culas, 3004-516~Coimbra, Portugal}\\
{$^{3}$Smoluchowski Institute of Physics, Jagiellonian University of Cracow, 30-059~Krak\'{o}w, Poland}\\
{$^{4}$Gesellschaft f\"{u}r Schwerionenforschung mbH, 64291~Darmstadt, Germany}\\
{$^{5}$Institut f\"{u}r Strahlenphysik, Forschungszentrum Dresden-Rossendorf, 01314~Dresden, Germany}\\
{$^{6}$Joint Institute of Nuclear Research, 141980~Dubna, Russia}\\
{$^{7}$Institut f\"{u}r Kernphysik, Johann Wolfgang Goethe-Universit\"{a}t, 60438 ~Frankfurt, Germany}\\
{$^{8}$II.Physikalisches Institut, Justus Liebig Universit\"{a}t Giessen, 35392~Giessen, Germany}\\
{$^{9}$Istituto Nazionale di Fisica Nucleare, Sezione di Milano, 20133~Milano, Italy}\\
{$^{10}$Institute for Nuclear Research, Russian Academy of Science, 117312~Moscow, Russia}\\
{$^{11}$Physik Department E12, Technische Universit\"{a}t M\"{u}nchen, 85748~M\"{u}nchen, Germany}\\
{$^{12}$Department of Physics, University of Cyprus, 1678~Nicosia, Cyprus}\\
{$^{13}$Institut de Physique Nucl\'{e}aire d'Orsay, CNRS/IN2P3, 91406~Orsay Cedex, France}\\
{$^{14}$Nuclear Physics Institute, Academy of Sciences of Czech Republic, 25068~Rez, Czech Republic}\\
{$^{15}$Departamento de F\'{\i}sica de Part\'{\i}culas, University of Santiago de Compostela, 15782~Santiago de Compostela, Spain}\\
{$^{16}$Instituto de F\'{\i}sica Corpuscular, Universidad de Valencia-CSIC, 46971~Valencia, Spain}\\

\end{raggedright}

 \newpage

\markboth{ {I.~Fr\"{o}hlich, J. Pietraszko \it et al.}}{Elementary Collisions with HADES}

\begin{center}
\textbf{Abstract}
\end{center}
The ``High Acceptance DiElectron Spectrometer'' (HADES) at GSI, Darmstadt, is
investigating the production of $e^+e^-$ pairs in $A+A$, $p+A$ and $N+N$
collisions. The latter program allows for the reconstruction of individual
sources. This strategy will be roughly outlined in this
contribution and preliminary $pp/pn$ data is shown.

\section{Introduction}

Recently, the HADES collaboration has reported on the production of di-leptons
- unique probes to study properties of dense hadronic
matter~\cite{Stroth:2006wh} - in the C+C collision at 2 AGeV~\cite{prl}. In
this context, the vector mesons are of particular interest as it has been
proposed that their spectral functions change inside the hadronic
medium~\cite{Brown:1991kk}. On the other hand, at these energies measured
di-lepton spectra do not contain the vector meson signal only, but also
additional lepton pairs from other sources, like $\Delta^{+,0}\to N\pi^{0}\to
N\gamma e^{+}e^{-}$ ($\pi$-Dalitz), $\Delta\to Ne^{+}e^{-}$ ($\Delta$-Dalitz),
$N^{*}(1535)\to N\eta\to N\gamma e^{+}e^{-}$ ($\eta$-Dalitz) and the decay of
baryonic $N^*$ resonances in $N(\omega,\rho)$.  This means at beam energies of
1-2 AGeV, corresponding to moderate densities (2-3 $\rho_0$) and temperatures
(60-80~MeV), the production of \mbox{(vector-)}mesons is {\it always}
accompanied by multi-step excitations of a limited number of resonances and
their subsequent decays, a concept which is supported many theoretical models
and also corroborated by recent HADES results~\cite{dohrmann,plb}.

One of the most abundant ingredients in this cocktail, the long-lived (i.e.
decaying after the freeze-out of the fireball) pseudoscalar mesons $\pi,\eta
\to \gamma \gamma^* \to \gamma e^{+}e^{-}$, have a well described
electromagnetic structure~\cite{landsberg}. Hence, they can be regarded as
``trivial'' components that can be subtracted from the measured $e^+e^-$
spectrum.  On the other hand, the contribution from short-lived resonances is
completely unknown. For example, the Dalitz decay of the $\Delta$ resonance
has not been measured. In the overall picture, these contributions are
additional exchange graphs in the virtual bremsstrahlung process $NN\to
NN\gamma^*$. One of the the questions recently addressed by one-boson exchange
models is how the resonance contributions have to be treated among with the
bremsstrahlung in coherent calculations, but a debate on this is still
ongoing~\cite{kaptari,mosel}.  The general conclusion is, however, that
experimentally a strong isospin dependence should be visible in the
mass-dependent ratio $M^{pp}_{ee}/M^{pn}_{ee}$.

To conclude this introduction, for an understanding of di-lepton spectra and
resolving medium effects from the vacuum spectral functions it is crucial to
fully describe the different sources in the heavy ion cocktail.  This means
that branching ratios and (spin dependent) form factors have to be known a
priori, as well as the features of the different production mechanisms as they
serve as input for model calculations since they affect mass and momentum
distributions.

\section{The elementary collision program}

In order to study these different processes, HADES (for a description
see~\cite{dohrmann}) has started a detailed program on the di-lepton production
in elementary collisions using a liquid hydrogen target and proton/deuteron
beams.  The first experimental run using a proton beam with a kinetic beam
energy of 2.2~GeV was successfully carried out in the year 2004 with the objective of
verifying the lepton pair reconstruction efficiency using the
known $\eta$ production parameters~\cite{if}. Moreover, $e^+e^-$ invariant mass results
can be compared to the $C+C$ experiment done at a similar kinetic energy per
nucleon (2~AGeV) thus providing an important reference.

\begin{figure}
\begin{center}
\resizebox{0.49\columnwidth}{!}{%
  \includegraphics{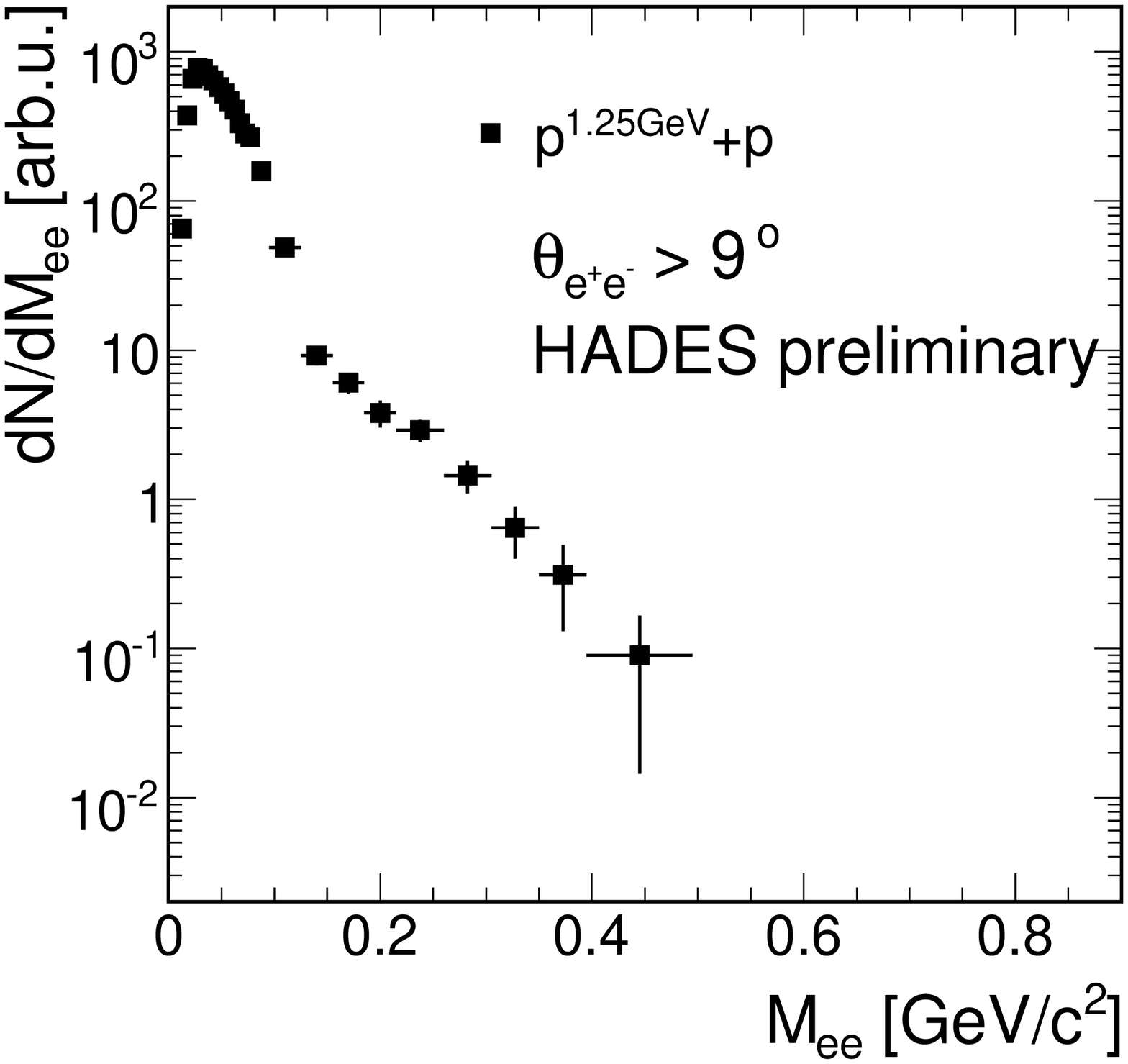}
}
\resizebox{0.49\columnwidth}{!}{%
  \includegraphics{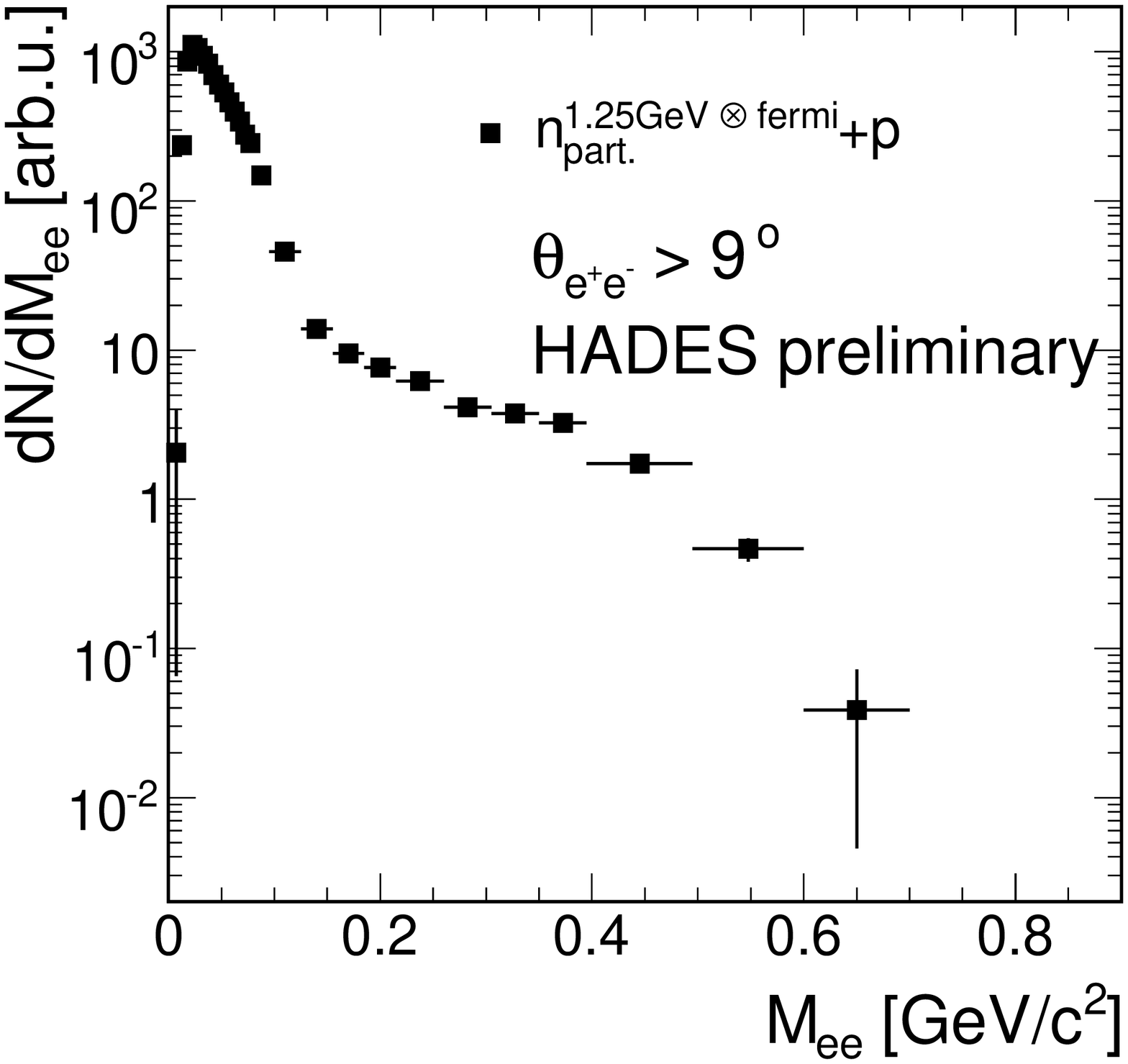}
}
\caption{The raw inclusive spectra obtained in the $pp$ reaction at
  1.25~GeV (left) and in the $np$ reaction (right, tagged with a proton
  spectator using the $dp$ reaction at 1.25~AGeV). 
  The
  spectra are not corrected for efficiency.
  In addition, for the $pn$ case the fermi motion has to be taken into
  account. Thus, the spectra cannot be compared directly.
  \newline ~}
\label{fig:pp_pn_p}       
\end{center}
\end{figure}

In the following, two experiments at a lower beam energy were performed: $pp$
at 1.25 GeV and $dp$ at 1.25 AGeV (N.B. the same energy per nucleon).  The
general idea of these consecutive measurements below the $\eta$ production
threshold was to focus on the low-mass continuum of the di-lepton cocktail.
By means of inclusive and exclusive analyses the $\Delta$ Dalitz decay can be
studied in the $pp$ case, as described below, where the bremsstrahlung is
negligible. The usage of iso-spin arguments allow for the subtraction of the
$\Delta$ contribution in the quasi-free $pn$ collision of $dp$ data set and to
study the virtual $pn$-bremsstrahlung.  This was made possible by the addition
of a forward wall which detected the spectator proton and thus tagged the
quasi-free $pn$ reaction. Fig.~\ref{fig:pp_pn_p} shows preliminary invariant
mass spectra for both the $pp$ as well as the $pn$ reaction. Here, the $pp$
results was obtained from $2.6 \cdot 10^9$ events, the $pn$ data stems from a
preliminary on-line analysis ($2.4 \cdot 10^9$ events, which is 50\% of the
expected statistics). Although the scales are
arbitrary, isospin effects are clearly visible in the mass region
above the $\pi^0$ peak, which points to additional sources like $pn$
bremsstrahlung and sub-threshold $\eta$ production.
This has to be discussed using detailed model calculations once the data set
has been finally analyzed.

It is clear that before a quantitative conclusion can be drawn a thorough
understanding of the $\Delta$ Dalitz decay (form factor and branching ratio)
is mandatory.  The production mechanism and decay of the $\Delta$ resonance
$\Delta \to N \pi$ is already known from older measurements~\cite{dmitriev,
  wicklund}. This fixes the $\Delta$ and $\pi^0$ emission angle and momentum,
which is modeled in the event generator Pluto~\cite{pluto}.  Thus, the
``trivial'' $\pi^0$ component can be subtracted and the $\Delta$ decay can be
studied by comparing several theoretical models with our data.

In the most recent experiment, $pp$ collisions at 3.5 GeV have been studied,
where $\omega$ and $\rho$ are produced with large cross sections. Here, the
focus of the analysis is to determine the contribution of vector production
mechanisms (partial waves and $N^*$ resonance contributions). The inclusive
$\omega$ line shape will serve as a reference for future $p+A$ reactions which
will be done at the same kinetic beam energy. The preliminary analysis of
on-line data suggests a yield of a few hundred  $\omega$-mesons in the
inclusive pair spectrum.  In addition, new precise data on the vector meson
production at higher beam energies are absolutely necessary, so the only data
on exclusive $\rho$,$\omega$ production come from old bubble chamber
experiments~\cite{bubble} with a yield of $\omega$ mesons of about 100 counts.
No differential cross section distributions have been reported in this energy
regime so far, but they are important inputs for the model calculations to be
done for the upcoming $p+A$ experiment.

\section{Summary and outlook}

In summary, HADES has successfully initiated a detailed program using
elementary collisions: The $pp$ and $pn$ reactions at 1.25~GeV to study
$\Delta$ production and decay, and $pp$ collisions at 2.2~GeV and 3.5~GeV. The
preliminary analysis of the data suggest a statistics large enough to clarify
many open question such as the $\Delta$ Dalitz decay, the $pn$ bremsstrahlung
and contributions from resonances. In particular, the preliminary analysis of
the $pp$ and $pn$ reactions seems to exhibit additional sources, which supports ongoing
theoretical work.

\section*{Acknowledgments}
{\footnotesize  The collaboration gratefully acknowledges the
support by BMBF grants 06MT238, 06GI146I,
06F-140, and 06DR135 (Germany), by the DFG
cluster of excellence Origin and Structure of
the Universe (www.universe-cluster.de), by GSI
(TM-FR1,GI/ME3,OF/STR), by grants GA CR
202/00/1668 and MSMT LC7050 (Czech Re-
public), by grant KBN 1P03B 056 29 (Poland),
by INFN (Italy), by CNRS/IN2P3 (France), by
grants MCYT FPA2000-2041-C02-02 and XUGA
PGID T02PXIC20605PN (Spain), by grant UCY-
10.3.11.12 (Cyprus), by INTAS grant 03-51-3208
and by EU contract RII3-CT-2004-506078.}





\begin{thebibliography}{0}    

\bibitem{Stroth:2006wh}
  J.~Stroth,
  J.\ Phys.\ Conf.\ Ser.\  {\bf 50} (2006) 389.

\bibitem{prl}   G.~Agakichiev {\it et al.},
  Phys.\ Rev.\ Lett.\  {\bf 98} (2007) 052302.

\bibitem{Brown:1991kk}
  G.~E.~Brown and M.~Rho,
  Phys.\ Rev.\ Lett.\  {\bf 66} (1991) 2720.

\bibitem{dohrmann}
  F.~Dohrmann {\it et al.},
  these proceedings.

\bibitem{plb} G.~Agakichiev {\it et al.},
  Phys.\ Lett. B, submitted [arXiv:0711.4281].



\bibitem{landsberg}
L.G. Landsberg,
Phys.Repts. \textbf{128}, (1985) 301.

\bibitem{kaptari}
  L.~P.~Kaptari and B.~K\"ampfer,
  Nucl.\ Phys.\  A {\bf 764} (2006) 338
  [arXiv:nucl-th/0504072].

\bibitem{mosel}
  R. Shyam and U. Mosel, Phys. Rev. {\bf C 67} (2003) 065202.

\bibitem{if}
  I.~Fr\"ohlich {\it et al.},
  Eur.\ Phys.\ J.\  A {\bf 31} (2007) 831.








\bibitem{dmitriev} V. Dmitriev, O. Sushkov, and C. Gaarde, Nucl. Phys. {
\bf A459} (1986) 503.

\bibitem{wicklund} A.B. Wicklund  {\em et al.}, Phys. Rev.  {\bf D 35} (1987) 2670.

\bibitem{pluto}  I.~Fr\"ohlich {\it et al.}, PoS (2007) ACAT 076 [arXiv:0708.2382].

\bibitem{bubble} A.P. Colleraine and U. Nauenberg, Phys. Rev. {\bf 161}, 1387 (1967). 
\\ L. Bodini {\em et~al.}, Nuovo Cimento {\bf 58A}, 475 (1968).



\end{thebibliography}
\end{document}